# QUENCH LIMITS IN THE NEXT GENERATION OF MAGNETS

E. Todesco, CERN, Geneva, Switzerland

*Abstract*

Several projects around the planet aim at building a new generation of superconducting magnets for particle accelerators, relying on $Nb_3Sn$ conductor, with peak fields in the range of 10-15 T. In this paper we give an overview of the main challenges for protecting this new generation of magnets. The cases of isolated short magnets, in which the energy can be extracted on an external dump resistor, and chain of long magnets, which have to absorb their stored energy and have to rely on quench heaters, are discussed. We show that this new generation of magnets can pose special challenges, related to both the large current density and to the energy densities.

## INTRODUCTION

Protection of superconducting magnet is a fascinating subject that involves different branches of physics and engineering, as material properties at low temperatures, superconductivity, heat propagation and magnet design. For the new generation of accelerator magnets, aiming at the 10-15 T range provided by $Nb_3Sn$, protection becomes a critical aspect.

It is usually stated that higher fields mean larger stored energies and this entails more challenging protection. This statement is not completely correct, since for long magnets the physical limit for hotspot temperatures is on the energy density in the coil rather than on the magnet stored energy. Indeed, this density in many $Nb_3Sn$ models is twice w.r.t. previous Nb-Ti accelerator magnets: there is no doubt that the new generation of magnets enters a new regime from the protection point of view: including protection from the start of the magnet design process is a must.

Here we will try to address the main issues in the interaction between magnet design and protection for accelerator superconducting magnets. We will give a special emphasis to the case of $Nb_3Sn$ conductor, which is being considered for the LHC upgrades. Starting with a discussion of the hotspot temperature, we outline the protection strategies with and without external dump, providing the relation to the main design parameters as current and inductance.

We then introduce the concept of time margin for protection, i.e. the time available to the protection system to quench all magnet before it reaches the limit in the hotspot temperature. We estimate this parameter for several cases, and we give the dependence on the design features, pointing out the relevance of the current density. The time margin is consumed by different operations of the protection system: we discuss here the detection time, related to the initial quench velocity, and the time needed by the heaters to quench the entire coil, which are two essential features of the problem. We conclude with a discussion of the inductive voltages which arise by an unbalance between parts of the magnet that are quenched and parts that are still superconductive.

## HOTSPOT TEMPERATURE

### Recall of adiabatic approximation

The basis of our analysis is the adiabatic equation of balance between heat given by Joule effect and specific heats

$$\int_0^\infty [I(t)]^2 dt = \nu A^2 \int_{T_0}^{T_\infty} \frac{c_p^{ave}(T)}{\rho_{Cu}(T)} dT \qquad (1)$$

where $I$ is the current in the magnet, $A$ the cross sectional surface of the cable, $\nu$ the fraction of copper, $c_p^{ave}$ the average volumetric specific heat, and $\rho_{Cu}$ the copper resistivity. Together with the Joule heating equation, one has a set of coupled nonlinear equations giving the current decay in the magnet $I(t)$, in the adiabatic approximation [1], and one can estimate the final temperature $T_\infty$ in the coil. Note that since the resistivity depends on the magnetic field, the final temperature also depends on the position in the coil.

The right hand side of (1), integrated up the maximum acceptable temperature $T_{max}$, is our "quench capital", i.e. what nature gives us to spend in terms of specific heats and resistivity to absorb the energy of the magnet:

$$\Gamma(T_{max}) \equiv \nu A^2 \int_{T_0}^{T_{max}} \frac{c_p^{ave}(T)}{\rho_{Cu}(T)} dT = \nu A^2 \int_{T_o}^{T_{max}} \gamma(T) dT ; \quad (2)$$

its physical units are square of current times seconds, usually expressed in MIITs. The quench capital $\Gamma$ depends only on the composition of the cable and on the magnetic field. It <u>scales with the square of the cross-sectional surface of the cable $A$, and is proportional to the copper fraction $\nu$</u>.

The left hand side of Eq. (1) is the "quench tax", i.e. what is consumed by the magnet

$$\Gamma_q \equiv \int_0^\infty [I_q(t)]^2 dt \qquad (3)$$

The quench tax depends on the features of the magnet as inductance, current, and on the circuit (energy extraction, etc). It <u>scales with the square of the current</u>.

### What to include in the capital

In the adiabatic approximation one has to make a hypothesis about the elementary cell that takes the heat. The most conservative hypothesis is to take the strand, i.e. the mix of superconductor and stabilizer. One can also assume that the Joule heating is also shared by the insulation and the epoxy (for impregnated coils). If the coil is not impregnated and operates in superfluid Helium, the contribution of HeII to the specific heat is very large below the transition temperature 2.17 K, so it plays a major role in the initial part of the heating. On the other hand, it becomes negligible w.r.t. the strands when the specific heat is integrated up to room temperature.

Elements which are more far from the original source of heating will take more time to contribute to the enthalpy. With typical time scales of the current discharge (0.1-0.5 s) the usual approximation takes into account of strand and insulation, but not wedges of the mechanical structure around the coil, see [2, 3] for more details.

In the following we will make the usual assumption of adiabatic codes, i.e., that the whole insulated coil shares the Joule heating, and the quantities in (1)-(3) will be referred to the insulated cable.

*Limits to hotspot temperature*

What is the maximum tolerable hotspot temperature guaranteeing no permanent degradation of magnet performances? A conservative limit can be established at 150 K [4, 5], and in most cases room temperature is considered to be safe. Some experiments on $Nb_3Sn$ magnets showed no degradation up to 400 K [3], and even more. For $Nb_3Sn$ magnets the temperature where the impregnation undergoes a phase transition can be considered a hard limit, see [3] for more details.

Since the quench capital $\Gamma$ approximately scales with the square root of the temperature [1], from 300 to 400 K one gets about 15% more, i.e. not such a dramatic increase. In the following, we will consider 300 K as a limit, knowing that this is a conservative value.

## PROTECTION STRATEGIES

*External dump resistor*

We first assume that the energy can be extracted to an external dump resistor $R_d$. The larger the dump resistance, the faster the decay:

$$I(t) = I_0 \exp\left(-\frac{t}{\tau}\right) \quad (4)$$

where, neglecting the magnet resistance, the time constant can be expressed as:

$$\tau = \frac{L}{R_d}, \quad (4)$$

The faster decay, the smaller is the quench tax $\Gamma_q$ (see Eq. 3). The limit to having large resistors is given by the voltage on the magnet

$$R_d < \frac{V_{max}}{I_o} \quad (5)$$

where the maximum voltage is of the order of 1 kV. Taking the maximum allowed limit, the quench tax $\Gamma_q$ is given by

$$\Gamma_q = \int_0^\infty [I(t)]^2 dt = I_o^2 \int_0^\infty \exp\left(-\frac{2t}{\tau}\right) dt = \frac{L}{R_d}\frac{I_o^2}{2} = \frac{UI_o}{V_{max}} \quad (6)$$

Where $U$ is the magnet energy (we assume the linear case with constant inductance). So the condition of protection reads

$$\Gamma > \Gamma_q = \frac{UI_o}{V_{max}}; \quad (7)$$

The first observation is that $\Gamma$ is independent of the magnet length, whereas $\Gamma_q$ is proportional to the length through the energy $U$. Therefore for longer and longer magnets the dump resistor strategy is less and less effective, due to the voltage limitation, i.e. the external dump resistor strategy is not independent of the magnet length. The second remark is that given an energy $U$, a magnet with larger cable (and less turns, i.e. lower inductance) has a more favourable energy extraction. In fact, the quench capital scales with the square of the area of the cable, whereas the quench tax scales with the current (see r.h.s. of Eq. 7), i.e. with the area of the cable. So in a case of external dump resistor, larger cable, higher currents and lower inductance ease protection, possibly allowing to satisfy the condition (7).

As an example, we show the case of the insertion quadrupole Q4 for the LHC upgrade. This magnet has to provide 550 T of integrated gradient, and is individually powered. Considering a two-layer coil with 8.8 mm width cable, one obtains 128 T/m operational gradient with 20% margin on the loadline. This option does not satisfy the quench protection requirement (7), i.e. the external dump resistor cannot provide a full protection (see Table 1). On the other hand, a one layer option with double width cable of 15 mm allows a protection with the dump resistor only as $\Gamma$ becomes greater than $\Gamma_q$.

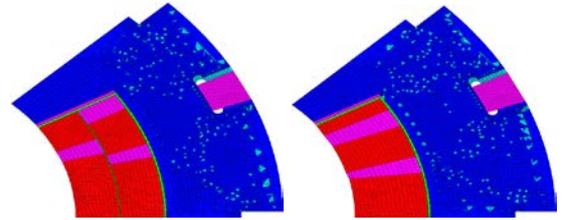

Figure 1: Cross-section of Q4 quadrupole for the LHC upgrade, one layer (left) and two layers (right) [6].

Table 1: Two options for the design of Q4 in LHC upgrade

|  |  | Two layers | One layer | Ratio |
|---|---|---|---|---|
| Integrated gradient | (T) | 544 | 544 | 1.00 |
| Gradient | (T/m) | 128 | 120 | 1.07 |
| Cable width | (mm) | 8.8 | 15.1 | 0.58 |
| Cable thickness | (mm) | 1.00 | 1.74 | 0.57 |
| Cable cross-section | (mm$^2$) | 8.96 | 26.11 | 0.34 |
| Length | (m) | 4.25 | 4.53 | 0.94 |
| Inductance | (H) | 0.086 | 0.0069 | 12.46 |
| Current | (A) | 4865 | 16188 | 0.30 |
| Dump resistor | (Ω) | 0.164 | 0.049 | 3.33 |
| Time constant | (s) | 0.523 | 0.140 | 3.75 |
| $\Gamma$ | (MIITs) | 3.2 | 30 | 0.11 |
| $\Gamma_q$ | (MIITs) | 6.2 | 18.4 | 0.34 |

*No external dump resistor*

If the dump resistor extracts only a negligible fraction of the stored energy, the magnet itself has to take this energy. Since the quench propagation is too slow (of the order of 1 s for a 10-m-long magnet), one makes use of quench heaters that induce a fast (of the order of 10-50 ms) transition to resistive state in most of the magnet.

The best that one can do is to evenly spread this energy all over the magnet, and a trivial limit is the balance between the energy to dissipate and the energy needed to bring the coil to $T_{max}$. This sets an intrinsic limit to protection without dump resistor. We therefore consider the integral of the specific heat of the coil

$$C_p^{ave} \equiv \int_{T_o}^{T_{max}} c_p^{ave}(t)dt \; ; \qquad (8)$$

this value depends on the materials (see Table 2). Nb-Ti has larger values than $Nb_3Sn$, but including insulation and voids (for Nb-Ti), they have a rather similar value of 0.5 J/mm$^3$. HTS materials have an enthalpy which is also in the same order of magnitude: here a reference cable is less established, and in Table 2 we give the YBCO tape used for FrescaII insert [7] and a BSSCO Rutherford cable.

Table 2: Typical integral of specific heat from 1.9 K to 300 K for superconducting coil used in accelerator magnets

|  | SC | | Stabilizer | | Insulation | | Total |
|---|---|---|---|---|---|---|---|
|  | (J/mm$^3$) | (%) | (J/mm$^3$) | (%) | (J/mm$^3$) | (%) | (J/mm$^3$) |
| Nb-Ti | 0.64 | 0.33 | 0.71 | 0.33 | 0.27 | 0.18 | 0.49 |
| Nb$_3$Sn | 0.46 | 0.33 | 0.71 | 0.33 | 0.27 | 0.33 | 0.48 |
| BSSCO | 0.55 | 0.18 | 0.57 | 0.53 | *0.27* | *0.18* | 0.44 |
| YBCO | negligible | | 0.65 | 0.67 | *0.27* | *0.33* | 0.53 |

Having found the hard limit, how far are we from it? An analysis of magnets built in the past 10-20 years, both short models and full-feature accelerator magnets as the LHC main dipoles, show that Nb-Ti magnets have an energy density of about 0.05 J/mm$^3$, i.e. about a factor 10 lower than the enthalpy limit. Since the energy scales with the square of the current, these magnets have about a factor three safety in current. For Nb$_3$Sn magnets we are entering a new regime, with energy densities of 0.10-0.12 J/mm$^3$, i.e., doubling the values of Nb-Ti and reducing the current margin from a factor 3 to a factor 2 only (see Fig. 2).

The capital Γ is independent of magnet length, and depends only on cable size. The tax in the case of no dump resistor is also independent of length, since both $R$ and $L$ scale with magnet length:

$$I_q = I_o \exp\left(-t\frac{R(t)}{L}\right); \qquad (9)$$

So the good news is that the <u>no dump strategy is independent of magnet length</u> – in this approximation, what works for a 1-m-long model will also work for 15-m-long magnet. Other phenomena, as quench-back, helping to spread the quench and increase the resistance may have a dependence on the magnet length.

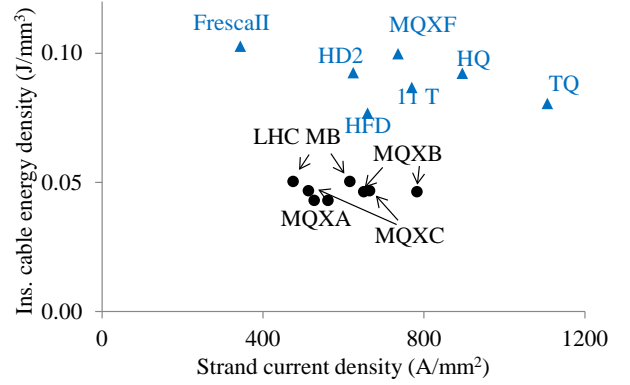

Figure 2: Energy density in the insulated coil for Nb-Ti (black circles), Nb$_3$Sn (blue triangles), and sketch of the limit given by the integral of specific heat.

## TIME MARGIN

*Definition*

Many margins have been defined for superconducting magnets, as the loadline margin, the temperature margin, and the current margin. Here we define a margin for protection in case of no dump resistor. We define the capital as in (2), and the tax as in (3), and we consider a discharge of the magnet shorted on itself, without dump resistor, and having the whole magnet quenched at $t=0$. This is defining an ideal protection system that is able to quench the entire magnet instantaneously - difficult to make something better.

Instead of defining the margin in terms of "fiscal pressure", i.e. the ratio between tax and capital, we see how much of what is left after tax can be spent staying at operational current (see Fig. 3).

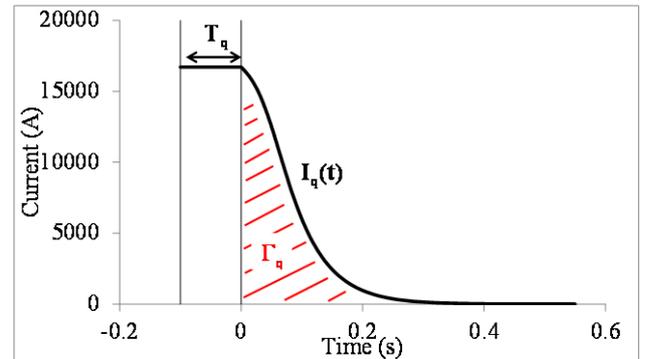

Figure 3: Definition of protection time margin.

$$I_0^2 T_q(T_{max}) + \Gamma_q = \Gamma(T_{max}) \qquad (6)$$

$$T_q(T_{max}) \equiv \frac{\Gamma(T_{max}) - \Gamma_q}{I_0^2} \qquad (7)$$

This is the time margin for protection, allowing to judge if the reaction time of the quench protection system is sufficient or not. The advantage is that the capital left is divided by the square of the current, allowing direct comparison between magnets with different currents (and therefore different cable cross-section, inductances, etc.).

*Scaling*

If we consider two cases as in Fig. 1, i.e. if we go from two to one layer, doubling the coil width, and keeping the same energy density, the time margin does not change. We remind the reader that to compute the time margin we always consider the case with a magnet fully quenched at time zero. In fact current will double, inductance will be divided by four, but resistance as well (a factor two from shorter cable, and a factor two from cross-section) so the time constant is preserved. So in case of no external dump, rearranging the same coil with less layers and larger cables does not affect the time margin.

*Dependence on magnet features*

An estimate of the time margin for several accelerator magnets is given in Fig. 4. One can see that Nb-Ti magnets as the LHC main dipole and the Nb-Ti option for the inner triplet upgrade have a time margin of 100-200 ms (depending on the layer, since the magnet is graded). Many cases of the new generation of $Nb_3Sn$ have a time margin reduced by more than a factor two, to about 50 ms (HD2, HFD, MQXF and 11 T). The 90 mm and 120 mm LARP quadrupole have even lower margin, 20-30 ms. It is interesting to see that the copper fraction which is usually considered to be the crucial parameter for protection, is not the only player in the game. An important variable is also the current density, which plays a major role. One can prove [8] that the time margin scales with the inverse of the square of the current density

$$T_q = \frac{v}{\bar{\rho} j_0^2} \left[ C_p^{ave} - \eta U_d \right] \quad (8)$$

and, moreover, it depends on some intensive properties of the magnet as the integral of specific heat as defined in (8) and the energy density $U_d$ over the coil, the copper fraction, and an average resistivity $\bar{\rho}$ of the stabilizer. Here $\eta$ is a parameter that depends on the magnet design (in our case in the range of 2-3 for most magnets) which hides the complexity of the problem.

This equation points out several interesting features:
- Provided that you manage to spread the quench in the whole magnet, relevant quantities for hotspot are intensive properties (energy density, current density, resistivity, copper fraction) and not extensive ones (energy, inductance, current). So large stored energies are not a problem for hot spot, but large energy densities are.
- There is a strong dependence on the current density, so an effective way of improving the aspects related hotspot temperature is to avoid too large current densities. This is clearly visible in Figure 5: FrescaII has a large time margin since its operational current density is very low (200 A/mm$^2$). Conversely, TQ is a particular difficult magnet to protect due to its large current density of 750 A/mm$^2$, giving only 20 ms of time to quench all magnet before reaching hotspot temperature.

The case of the LARP quadrupoles of 90 mm (TQ), 120 mm (HQ) and 150 mm aperture (MQXF) [9] is particularly interesting since without any change in copper ratio but reducing the current density one has obtained an easier protection, notwithstanding the increasing stored energy.

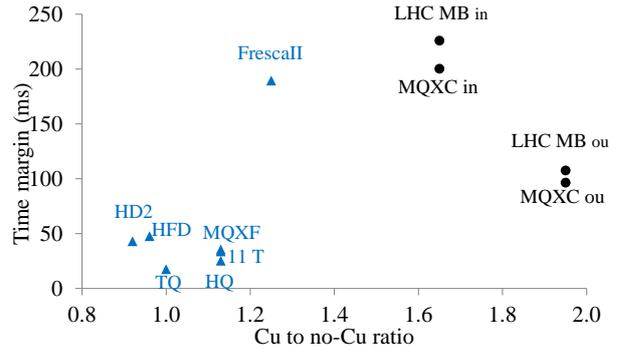

Figure 4: Time margin versus copper-no copper ratio, for Nb-Ti (black circles), and $Nb_3Sn$ (blue triangles) magnets.

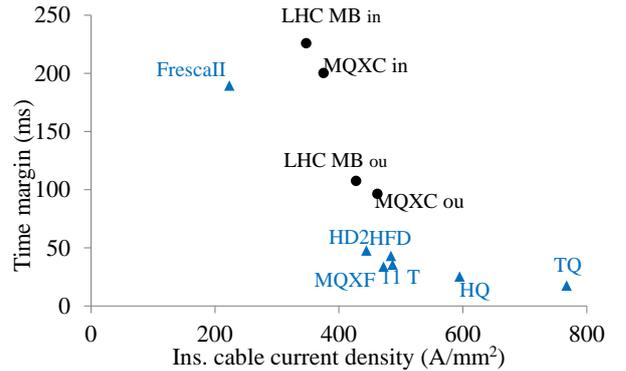

Figure 5: Time margin versus current density for Nb-Ti (black circles), and $Nb_3Sn$ (blue triangles) magnets.

*Dependence on local field*

At low temperatures, the resistivity of copper strongly depends on the magnetic field. At 1.9 K and 12 T the resistivity is five times lower than at 1.9 K and 0 T. For this reason if the heat does not get averaged over the whole coil, one can have large differences in the time margin between high field and low field zones.

As an example in Fig. 6 we consider the case of the 120 mm aperture quadrupole HQ: the time margin varies between 25 ms at 12 T up to 45 ms at 2 T. Note that since the cable is considered to be the basic cell, and due to the transposition of the cable, there is no strand in the magnet that sees less than 2 T. If there is no heat diffusion we will have a larger budget for the low field zones, which will become useful soon. If the heat diffusion plays an important role, this difference is smeared, with the effect of increasing the budget for the high field zones.

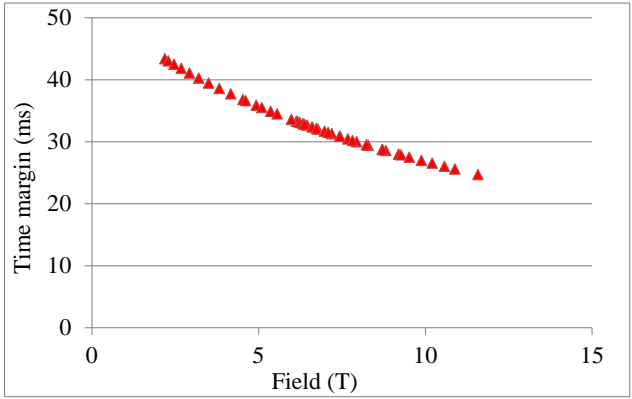

Figure 5: HQ time margin versus local field.

## HOW TO SPEND THE TIME MARGIN

### General budget

The time budget is the time available to the system to quench the entire magnet. It must be larger than the sum of different contributions:
- Detection time, i.e. the time needed to detect the quench. This is given by the resistance growth along the cable, related to the quench velocity, and to the thresholds used for detection.
- Validation time: once the threshold is reached there is a validation window (typically 5 to 10 ms) to avoid false triggers.
- Switch opening: typically 2 ms.
- Quench heater delay: time needed by the heaters to quench the magnet. It can be separated into:
  o Delay needed to start a quench somewhere in the coil – typically in the high field zone - what is usually measured as quench heater delay.
  o Delay between the start of the quench in the high field zone and the quench of the whole coil, including the low field zone.

In the following we will treat the detection time and the heater delay.

### Detection time and quench velocity

A quench is detected through the measurement of a voltage generated by the resistive transition. To compensate for the inductive voltage during the ramp, voltages of two symmetric poles are subtracted. The lower the threshold, the smaller is the detection time: typical values of thresholds used in the LHC are $V_{th}$=100 mV, fixed by the accelerator system.

The resistive voltage is proportional to the current and to the growing resistance, in first approximation as

$$V(t) = R(t)I_o = \frac{v_q t \rho}{A_{Cu}} I_o \qquad (9)$$

where we define the quench velocity $v_q$. The resistivity is constant within 10% in the range between 2 and 40 K, so one can write the detection time as

$$t_{det} = \frac{V_{th}}{v_q \rho} \frac{1}{j_{o,Cu}} \qquad (10)$$

Typical values of current density in the copper of 1400 A/mm$^2$, copper resistivity at 12 T of 0.65 nΩ m, and quench velocity of 20 m/s give a detection time of the order of 5 ms for a quench in a high field zone in a 12 T operational field Nb$_3$Sn magnet.

For low fields one has two negative effects that tend to increase the detection time:
- The resistivity is considerably lower for lower fields: at 2 T one has a resistivity of 0.22 nΩ m, i.e. a factor three lower.
- The quench velocity is lower for low fields: it is proportional to the square root of the conductivity times resistivity divided by the temperature margin:

$$v_q \propto \sqrt{\frac{\rho \kappa}{T_{cs} - T_o}} \qquad (11)$$

The temperature margin at 80% on the loadline, for Nb$_3$Sn is ~4.5 K in high field and ~12.5 K in low field (2 T), so the ratio is about a factor 3. The product resistivity times conductivity changes from 12 T to 2 T of a factor that can be estimated between 1 and 2, according to the sources. So in the most pessimistic case the quench velocity is 2.5 times smaller.

Putting together all the effects, one has a detection time which is 5 to 7 times larger in low field regions, increasing the detection time from 5 ms to 25 to 35 ms. These additional 20 to 30 ms are barely compensated by the larger budget available in the low field zone (see Figure 5). So quenches in low field zones are a critical issue due to larger detection times.

### Heater delay

With a longitudinal speed of 10-20 m/s, a typical propagation from turn to turn of ~10 ms and between layers of ~50 ms, it appears clear that the growth of resistance due to the quench propagation is negligible, and the only way to have a fast dump is to quench most of the magnet rapidly through the quench heaters. So, the core of the protection problem is to model how the heat of the quench heaters propagates to the coil and how long it takes to quench the different zones of the coil.

A simple model is based on the estimate of the energy needed to bring the coil from the operational temperature $T_o$ to the current sharing temperature $T_{cs}$

$$t_{del} \propto \int_{T_o}^{T_{cs}} c_p^{ave}(T) dT \qquad (12)$$

This model has one free parameter and allows to estimate the ratio in the delay between different conditions.

The geometry of the heaters can be rather complex [11, 12]. For long magnets, heaters covering completely the coil and providing the needed power would lead to too large voltages. Two strategies can be used to cope with this problem:

- heaters of smaller width, with a wavy shape that cover all cables every given longitudinal period (as in HQ);
- heaters with variable width, i.e. having heating stations spaces along the magnet axis (as in LQ). This can be also obtained with copper cladding as in the main LHC dipoles [13].

In both cases the magnet is quenched only in a few positions along the axis, and in between them one relies on quench propagation. For this reason the spacing of the stations or the period of the waves must be of the order of 100 mm, so that propagation over 50 mm at 20 m/s takes a few ms. A code is being recently developed to model the heat transfer from the heaters to the cables, relying on a 2 D thermal network, and allowing to simulate these complex geometries and to optimize them [12].

The delay has a nonlinear dependence on the current, being obviously zero at short sample, is roughly proportional to the thickness of the insulation between the heaters and the coil, and saturates at towards 50 W/cm$^2$ in case of a 0.025 mm thickness heater strip [14]. With the smallest insulation of 0.025 mm, and optimized heater power one can get, at 80% of the loadline, delays of the order of 5 ms. The delay, which is routinely measured during a test campaign, is obviously related to the quench in the higher field zone.

A second element that cannot be measured directly is the time needed to quench the lower field zones. The simple model based on the integral of the specific heat gives, again for the case of HQ, a factor 2.5 between the time needed to quench low field zone w.r.t. high field (see Figure 6). So in the hypothesis of 5 ms delay, another 10 ms are needed to completely quench the outer layer.

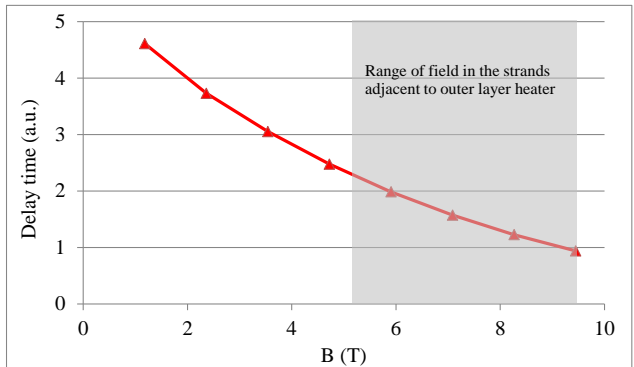

Figure 7: Delay time of quench heaters vs field for HQ.

The last piece of the puzzle is how to quench the inner layer. There are three possible strategies:

- have a quench heater glued on the inner part of the inner layer (as done in HQ). This heater, easy to add during the coil potting, has the disadvantage of not being supported and is prone to detachment that can reduce its efficiency during the magnet lifetime. Moreover, it constituted a barrier to the heat removal, which for this magnet is at 80% from the inner cold bore;
- have a quench heater in the interlayer. This option can be realized through building a heater that can resist to the Nb$_3$Sn heat treatment, or having a splice between the inner and the outer layer. The first option was tried and then abandoned in the HFD program [13];
- rely on the propagation of the heat from the outer to the inner layer, i.e. having the outer layer acting as a quench heater. This induces a considerable delay, which for the 11 T magnet has been estimated to 50 ms.

For both the MQXF, the inner triplet quadrupole of the LHC upgrade, and the 11 T dipole to make space for collimators, quenching the inner layer is a critical issue that is still to be solved. Putting together all the components, it appears clear that a time margin of 50 ms is at the limit of protection, whereas 20-30 ms are out of reach.

## INDUCTIVE VOLTAGES

During the discharge, if the magnet has no external dump resistor and it is individually powered or has a by pass diode, the inductive voltage compensates the resistive voltage, i.e.

$$L\frac{dI(t)}{dt} + R(t)I(t) = 0 \qquad (13)$$

where $L$ and $R$ refer to the induction and to the resistance of the whole magnet. If the magnet is quenched in the segment $a$ to $b$, and is superconductive in the segment from $b$ to $c$, with $a$ and $c$ being the ends of the magnet, one has

$$(L_{ab} + L_{bc})\frac{dI(t)}{dt} + R_{ab}(t)I(t) = 0 \qquad (14)$$

and a voltage appears within the magnet

$$V_{ab} = L_{ab}\frac{dI(t)}{dt} + R_{ab}(t)I(t) = -L_{bc}\frac{dI(t)}{dt} \qquad (15)$$

Please note that here $L_{ab}$ denote the inductance of a segment of the magnet, defined as the ratio of the measured inductive voltage during a discharge and the derivative of the current. The maximum acceptable inductive voltage is related to the magnet insulation and is of the order of 1 kV. One can point out the two main scaling related to inductive voltages:

- The <u>inductive voltage is proportional to the magnet length</u>. So for a given cross-section there is a maximum magnet length above which inductive voltages are not acceptable;
- <u>A magnet with larger cable has lower inductive voltages</u>; let us compare as for the hotspot temperature two cases with same field, stored energy, one with a double layer with width $w$, and the second with a single layer with width $w'=2w$. One has

$$I_o' = 2I_o \qquad L' = \frac{L}{4} \qquad R' = \frac{R}{4} \qquad (16)$$

and the voltage of the one layer case is a factor two smaller than the two layer case

$$V' = \frac{V}{2}. \qquad (17)$$

We estimated the inductive voltage in the case of a magnet which has the outer layer totally quenched and the inner layer superconductive. This is what happens at the beginning of the quench for magnets with outer layer quench heaters. We considered an extreme case where the inner layer never quenches. Simulations are shown in Fig. 7 shows for six magnets including the LHC main dipoles.

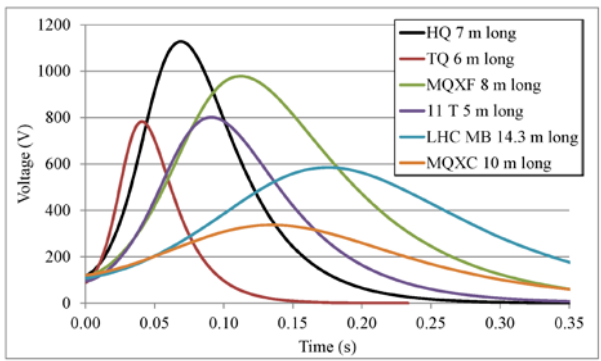

Figure 7: Estimate of inductive voltage in four $Nb_3Sn$ magnets, and in two Nb-Ti magnets, due to unbalance with inner layer not quenched and outer layer totally quenched.

In many cases the resistance of the outer layer is not enough to guarantee a hotspot temperature below 300 K: after a certain time one has to quench also the inner layer, so the significant part is this simulation are only the first 50-100 ms. It turns out that all cases are in the range of 500-1000 V, so close to the threshold. This means that these designs associated with their magnet lengths are just at the limit of the tolerance for this case: magnets with significantly longer length would require a different design of the coil.

## CONCLUSIONS

The new generation of $Nb_3Sn$ magnets, with peak fields in the range of 10-15 T, poses novel challenges for protection. Here we reviewed the aspects related to the hot spot temperature. We first considered the case of short magnets individually powered, which can be protected with external dump resistor and which profit of large cables and small inductances. Then we analysed the case of long magnets, or chain of short magnets, where the dump resistor strategy is not effective: large or small cables make no difference, and the magnet itself has to absorb its energy, relying on quench heaters.

We defined a novel concept of time margin, which gives the time allowed to the protection system to react before the magnet reaches too high temperatures, and we presented the relevant scaling law. This time margin allows to compare directly different designs and technologies. It turns out that if the Nb-Ti magnets had a time margin of 200-100 ms, with the new generation of $Nb_3Sn$ magnets the margin is 50 ms, and even 20-30 ms in some cases. The exception is FrescaII, since it relies on a very large coil and small current density.

We then discussed the time of reaction of the system, from the start of the quench to the instant at which all the magnet is quenched by the heaters. We discussed the different contributions, pointing out that a quench in a low field region can be as challenging as a quench in the high field regions, due to a larger margin, longer time to detect and a lower quench velocity. In general, one can state that 50 ms reaction time is challenging but typically achievable, whereas 30 or 20 ms seem impossible to achieve with present experience and technologies.

We finally discussed the main scaling for the inductive voltages that arise during quench. In this case, large cable and small inductances allow reducing the voltages. On the other hand, the voltages scale with the magnet length so they impose an upper limit to the magnet length.